\documentclass[referee,doublespacing]{gji}
\usepackage{timet,color}
\usepackage{graphicx}
\usepackage{xcolor}
\usepackage{soul}
\usepackage{amsfonts}
\usepackage[urlcolor=blue,citecolor=blue,linkcolor=black]{hyperref}
\usepackage{natbib}

\newcommand{\stkout}[1]{\ifmmode\text{\sout{\ensuremath{#1}}}\else\sout{#1}\fi}
\usepackage{indentfirst}
\usepackage[normalem]{ulem}

\title[Deep-tomography]{Deep-tomography: iterative velocity model building with deep learning}
\author[A.~Muller et al.]{Ana P.~O.~Muller$^1$ $^2$, Clecio R.~Bom$^2$ $^5$, Jessé C.~Costa$^3$ $^4$, Matheus Klatt$^2$, \and Elisangela L.~Faria$^2$, Bruno dos Santos Silva$^3$, Marcelo P. de Albuquerque$^2$, \and Marcio P. de Albuquerque$^2$\\
$^1$ Petróleo Brasileiro S.A. (PETROBRAS), Edifício Senado, Av. Henrique Valadares, 28, Centro,\\ Rio de Janeiro, Brazil. \\
$^2$ Centro Brasileiro de Pesquisas Físicas (CBPF), Rua Xavier Sigaud, 150, Urca, Rio de Janeiro, Brasil. \\
$^3$ Universidade Federal do Pará, Belém, PA, Brasil.\\
$^4$National Institute of Petroleum Geophysics (INCT-GP).\\
$^5$Centro Federal de Educação Tecnológica Celso Suckow da Fonseca (CEFET-RJ), Av. Maracanã, 229,\\ Maracanã, Rio de Janeiro, Brasil.}
        
\date{}
\pagerange{xx--xx}
\volume{xx}
\pubyear{xxx}

\begin{document}
\label{firstpage}

\maketitle
\begin{summary}
The accurate and fast estimation of velocity models is crucial in seismic imaging.
Conventional methods, like Tomography, Stereotomography, Migration Velocity Analysis (MVA) and Full-Waveform Inversion (FWI), obtain appropriate velocity models; however, they require intense and specialized human supervision and consume much time and computational resources.  
In recent years, some works investigated Deep Learning (DL) algorithms to obtain the velocity model directly from shots or migrated angle panels, obtaining encouraging predictions of synthetic models. 
This paper proposes a new flow to to recover structurally complex velocity models with DL. Inspired by the conventional geophysical velocity model building methods, instead of predicting the entire model in one step, we predict the velocity model iteratively. 
We implement the iterative nature of the process when, at each iteration, we train the DL algorithm to determine the velocity model with a certain level of precision/resolution for the next iteration; we name this process as Deep-Tomography. 
Starting from an initial model, that is an ultra-smooth version of the true model, Deep-Tomography is able to predict an appropriate final model, even in complete unseen during the training data, like the Marmousi model. When used as the initial model for FWI, the models estimated by Deep-tomography can also improve substantially the final results obtained with FWI.
\end{summary}


\section{Introduction}

\indent Velocity model building(VMB) is an important problem in exploration geophysics. The accuracy of the velocity model used in migration algorithms is responsible for the correct imaging and focusing of the subsurface structures, which directly affects the exploration success rate. 
Among the most used and investigated methods we can mention Tomography \cite[]{tomo,RAYTOMO}, Stereotomography \cite[]{stereo01,stereo02,stereo03}, Migration Velocity Analysis (MVA) \cite[]{MVA1,MVA2} and Full-Wave Inversion (FWI) \cite[]{tarantola,pratt}. Based on inverse problem
methods, they estimate the model iteratively and depend on expert human curating and computational
capacity to obtain an accurate result.

MVA methods, which use as input the migration results to update the velocity model, attract great research interest since they work with the seismic data in the same domain as the velocity model. The model update can be ray-based, also called tomographic methods \cite[]{Stork,tomo10} or based on the wave-equation \cite[]{WEMVA}.
Tomography applies linear iterations to solve a nonlinear problem. To mitigate the instability of the inverse method, the iterations start from a low-resolution model, usually  a heavily smoothed version of the interval velocity obtained from the Dix formula \cite[]{DIX} or other technique that produces a low-resolution approximation of the actual model \cite[e.g.,][]{init01}. Then, velocity errors are measured over the migrated data and are backpropagated by the Tomography, defining a new model. This new model presents fewer velocity errors when compared with its predecessor and is the input to a new iteration. The next iteration starts migrating the seismic data with the model obtained in the previous iteration. Then, successive iterations are applied to the model, increasing its accuracy and improving the migrated image \cite[]{tomotut,tomo10}.

In recent years, machine learning (ML), especially deep learning (DL) techniques, started to be intensively explored in geophysical problems for a wide variety of scales, instruments and data \cite[see, for instance, ][]{reviewGeoML,bom2021bayesian,dias2020automatic}, such as automatic detection of faults in seismic images \cite[]{araya-fault}, automatic detection of scattering points in the migration dip angle domain \cite[]{scater}, automatic detection of salt bodies in seismic images \cite[]{SALTdetection,SALTdetection2}, seismic image denoising and resolution enhancement \cite[]{DENOISE,DENOISE2} and, seismic facies segmentation \cite[]{SEMANTICSEISMIC}. One particular set of problems that attracted attention is the one related to VMB \cite[]{reviewMLvel}, from automated NMO velocity picking \cite[]{NMODL2,NMODL1} to FWI assisted by ML algorithms \cite[]{FWIDL1,FWIDL2,fwiDL02}, until the complete model estimation \cite[]{biswas,ARAYAPOLO1,ARAYAPOLO2,YANGBASE,shucai20,fabien,paper4,Fomel2021}.

Some works used DL to infer the subsurface model \cite[]{ARAYAPOLO2,YANGBASE,shucai20}, or changes in velocity model \cite[]{yuan}, directly from the shots acquisition. They are based on the assumption that the shots carry enough subsurface information to perform the task. However, using the shots directly generates problems like lack of correspondence between time events (shots) and depth events (models), unreal geometries used in the experiments where the receivers cover all the surface, besides and prohibitive large data size when we consider a real seismic acquisition. Moreover, even if these problems would be possible to overcome, fully determining the velocity model from the shots is a highly complex task, which needs constraints due to the ambiguities generated by the dual relation between the medium's velocity and position of the reflectors. Thus, such a task is likely to require DL models with high complexity, i.e., with huge numbers of parameters, nontrivial structures and extended training data-set to represent a great variety of possible events \cite[]{DEEPBOOK}.

A new approach that bypasses some problems mentioned with the prediction of velocity model from the shots and enables the use of a non-regular receivers distribution was independently proposed by \cite{Zhang22} and \cite{Fomel2021}. The approach consists in using a well-known source of velocity information in geophysics: the common image gathers (CIG) in the angle domain. 
Based on the idea that CIG information is long used in Tomography algorithms to detect the move-out of events and then correct the velocity model, the shots were migrated with a constant velocity to generate common-angle panels. Then, the common-angle panels serve as input to the DL algorithm, trained to predict the accurate velocity model. The proposed method presents exciting results, recovering structurally simple synthetic models with high accuracy.

In this paper, we also used a migration method to generate the input for the DL velocity inversion algorithm. However, instead of using the CIG in the angle domain, we use the subsurface offset.
This gather is simpler to generate and represents the velocity error information with fewer traces when compared with the angle domain. The idea of using this simpler gather is based on the fact that the DL algorithm infers information from the input information used for the training. Once the training is made with the subsurface offset, it does not require any extra transformation from angle gathers for a DL algorithm to infer velocity errors from the subsurface offset. Furthermore, considering that the traces of CIG are the channels of the images for the DL algorithm, the reduced number of traces of the subsurface offset domain defines a network with a reduced number of trainable parameters compared with the angle gather based input. 

One important innovation point, based on the conventional process of VMB, is that our method divides the velocity estimation process into steps, starting from a smooth initial model until reaching the accurate velocity model, with high resolution and complexity. We implemented the iterative process, defining that each iteration has its own specialised trained network, whose inputs are common-offset subsurface panels migrated with the model of the current iteration and whose labels (output) are the updating to be summed at the model used in migration. Due to the similarity of the inputs, and the same iterative nature of tomography methods, we called the process proposed in this work of Deep-Tomography. 
In order to proof the accuracy and efficiency of the method, we generated a set of synthetics velocity models with complex structures of faults, folds, and velocity inversions, which the DL successfully predicted. We also performed a test over the Marmousi model, obtaining encouraging results. 

\section{Data-set construction}

One problem when training a DL algorithm to predict geophysical data is the lack of labelled and trustful data. In order to have enough and diverse data for training, we generated a set of $1000$ marine models with $3.5$ km of depth and $9$ km of lateral extension, using a sequence of geological processes as described in the next section. 
For each model, we simulated a synthetic seismic acquisition by implementing the acoustic, isotropic wave equation as a finite-difference scheme \cite{finitediff} with second-order in time and eighth-order in space (with spatial sampling equal to 10~m), with CPML boundaries \cite[]{MartinPML,KOMASTISCHPML}.

We used a Ricker wavelet with a peak frequency of 20 Hz to simulate the source. The shots are recorded by 701 surface receivers with a distance of 10~m between them, in a split spread acquisition with the maximum offset relative to the source of 3.5~km. We recorded 225 shots. The first shot occurs at the beginning of the interest region and the last shot at the end, with 40~m between shots. We used a time sampling interval of 4 ms for a total recording time of 3.5 s. We applied the mute of direct wave in shots for migration. The models were laterally extrapolated to allow the full shot over the region of interest. 
\subsection{Geological folded/faulted models}
\label{sec:models}
\begin{figure*}
    \centering
    \includegraphics[scale=0.30,trim={1cm 1cm 1cm 1cm}]{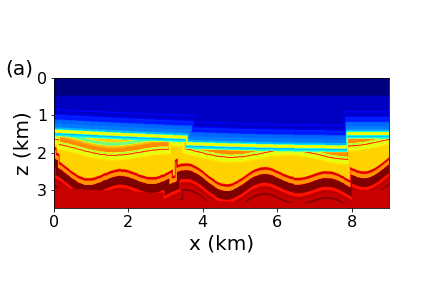}
    \includegraphics[scale=0.30,trim={1cm 1cm 1cm 1cm}]{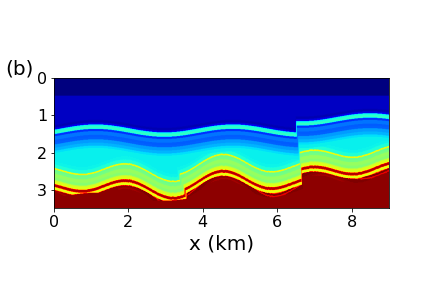}
    \includegraphics[scale=0.30,trim={1cm 1cm 1cm 1cm}]{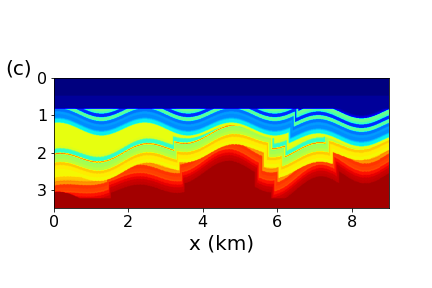}
    \includegraphics[scale=0.312,trim={1cm 1cm 1cm 1cm}]{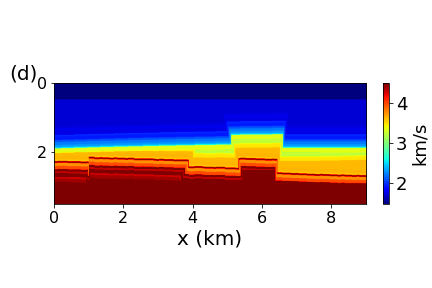}
    \caption{Figures (a), (b), (c) and (d) are some selected examples of models generated by the procedure described in Section \ref{sec:models}, where it is possible to observe the diversity of geometries generated, with faults and folds to increase the complexity of the models to be reconstructed by the proposed methodology.}
    \label{fig:models}
\end{figure*}

DL-based methods learnability is correlated to the variety and representativeness of the training set. Therefore, since the geological scenarios change drastically from one sedimentary basin to another, the best way to create a set model which represents the degree of complexity we aim to predict is an essential subject of study of VMB flows based on DL \cite[]{velnovo,velmod,velmod2}.

To make our synthetic models with high structural complexity, we simulated a system of deposition, folding, faulting, and erosion,  which aims to be a simplified version of the actual geological process. We made data simulations and model predictions over a 2-D mesh, with each 2-D model generated independently. The deposition process occurs in packages where a sequence of 12 deposited layers configures a package. The algorithm randomly chooses the thickest of each layer from a range of possible values. After the complete deposition of a package, we apply a folding process that distorts the layers by a sinusoidal function whose amplitude, phase, and period are random.
After the folding, we faulted the deposited layers, using a random number of normal faults with random angles and displacements.
Finally, an erosion process occurs with a fifty per cent probability, removing the randomly determined thickness of the previously deposited layers. 

The deposition, folding, faulting, and erosion processes are repeated three times, forming three packages. It is essential to mention that the previously deposited packages are subjected to the deformations of the packages above.
We adjusted the first layer, i.e., the water layer, to have a constant thickness of 0.5km, besides we defined a minimal distance of 0.25km from the sea bottom to the first sediment interface. 
After defining the layer's geometries, we define its velocities values. The velocity of each layer is chosen with values between 1.5km/s and 4.7km/s  meters, increasing with depth but allowing local velocity inversion. Layers with high velocity compared with their surroundings are also allowed to occur with thirty per cent probability. Fig.~\ref{fig:models} presents some examples of the variety of models that we generated for this study. We selected some examples to show that the random nature of the process produces a high diversity of models and velocity distributions. For example, the model in Fig.~\ref{fig:models}(d) is dominated by a fault regime, Fig.~\ref{fig:models}(b) shows a model where the most remarkable features are the folds, and models in Fig.~\ref{fig:models}(a) and (c) present examples of high complexity, where besides the faults and folds, an erosion process can also be observed.

\subsection{Migration of the common subsurface offsets}
We migrated the the recorded data using Reverse Time Migration~(RTM) with a cross-correlation extended imaging condition \cite[]{extIC}. RTM uses the full wave equation to back propagate the fields of the receivers $W_{r}(\mathbf{x},t)$ forward propagate the field of the source $W_{s}(\mathbf{x},t)$ into the earth, where $\mathbf{x}=\{x,z\}$ for our 2-D case. The generalized form can be written as:

\begin{equation}
R(\mathbf{x},\mathbf{\lambda},\tau)=\sum_{shots} W_{s}(\mathbf{x}-\mathbf{\lambda},t-\tau)W_{r}(\mathbf{x}+\mathbf{\lambda},t+\tau).
\label{eq:ic}
\end{equation}

Equation \ref{eq:ic} opens the possibility to vary the time $\tau$ and the space lag $\lambda$. In this work, we chose, for simplicity, to use only the horizontal spatial extension by applying $\mathbf{\lambda}$ only over the $x$ direction. $\lambda$ is known as the subsurface offset.

A gather presenting the subsurface offset, migrated with the correct velocity model, concentrates the reflections' energy in $\lambda=0$, correlations at non-zero subsurface offset indicate that the velocity model is incorrect. The focus of energy in zero subsurface offset occurs because the source and receiver fields correlate better at the correct reflection or scattering position when using the suitable model. The subsurface offset gathers were previously used in VMB problems \cite[e.g.,][]{suboff,TWI:IMAGE}; however, usually, the gather is converted to angle domain, and these angle gathers are then used in the flow of VMB \cite[]{angle}. 
In the DL context, the subsurface offset panels have a practical advantage over the angle domain, since the information is more compact, being necessary a reduced number of traces in the subsurface offset domain to represent the velocity error and events, reducing the size of the input data to the network training and prediction process.

\begin{figure*}
    \centering
    \includegraphics[scale=0.33]{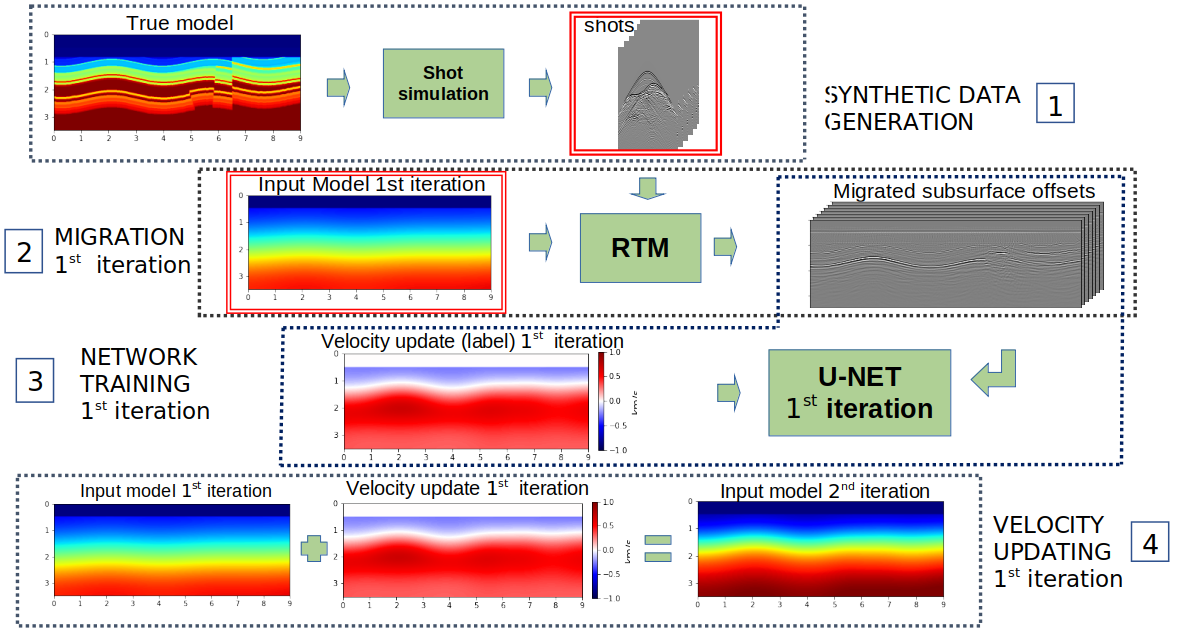}
    \includegraphics[scale=0.33]{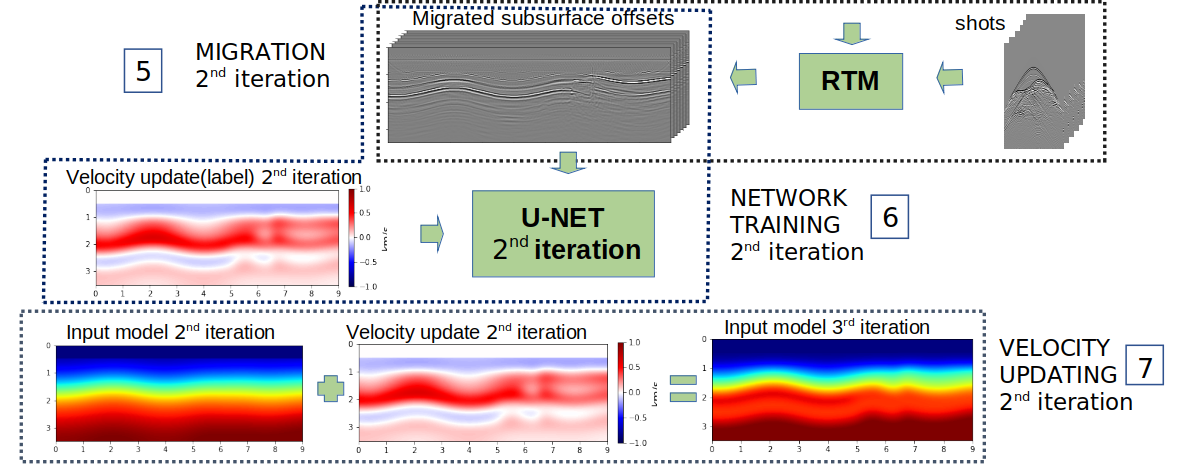}
    \includegraphics[scale=0.33]{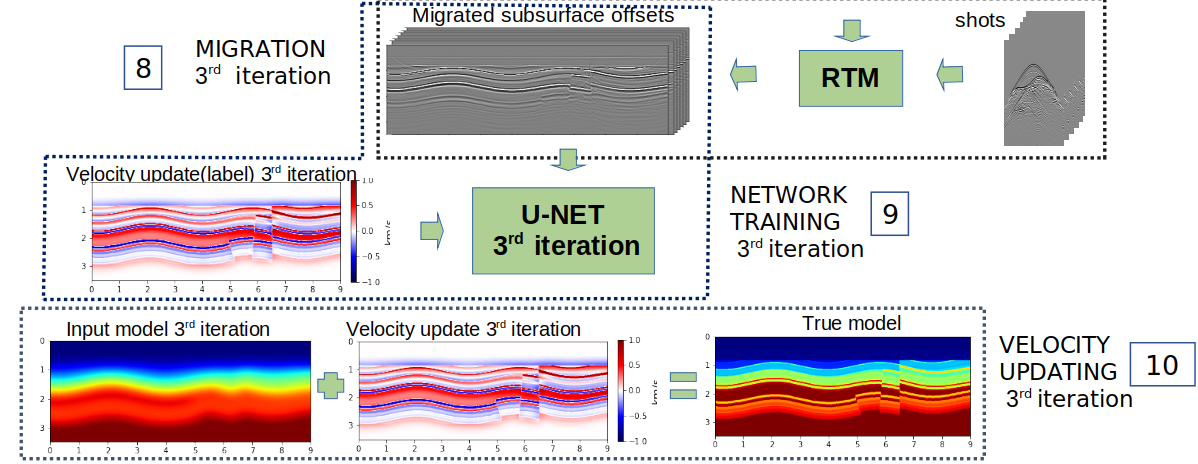}
    \caption{Complete Deep-Tomography flow. This figure represents a pipeline for training the three U-Nets. The labels, i.e., the desired output from the predictions, determine the next velocity reached in the flow. Once we finalized the training process, to obtain an unknown velocity model, it is only necessary to have as input the shots and the input model of first iteration, and the final result would be a detailed and accurate true model.}
    \label{fig:tomoflow}
\end{figure*}

\subsection{Defining the output model of each iteration}

A classic tomographic flow refines the velocity model iteratively. First, the seismic shots are migrated with a low frequency approximation of the actual velocity model, then the inaccuracies of such model are measured, usually by the residual move-out information, and back-propagated into the model. Then a new model is obtained and used to migrate the shots for a new iteration, which defines a new model update. The accuracy of the velocity model is increased at each iteration, usually manipulating variables like maximal resolution and maximal depth of interest until reaching the desired model accuracy. 

Assuming that it is challenging to predict in one step a complex velocity model like the ones generated for this study (see Fig.~\ref{fig:models}), we propose applying the same concept of successive iterations used in tomography with DL algorithms, to reach an accurate final model. We proposed that the DL algorithm for each iteration must be specialized to convert the migrated subsurface offset panels to the appropriated model update for the current iteration. Our DL algorithm will solve a supervised regression task, with subsurface offset panels as input, and the desired model update as output labels of the network.
Empirically, we chose to make three iterations of model updating. Initially, we tried to recover the entire model in one step, which generated a degraded and inaccurate model. Probably the migration algorithm destroyed the details of the high-frequency structures since they are not remotely represented in the model, and without information, the U-Net could not learn the suitable transformations. Then we introduced intermediary steps to our train/prediction process until we found the proposed configuration. The intermediary steps of model updating using different scales of smoothing is represented in Fig.~\ref{fig:tomoflow}. 

An essential step during a tomography is the choice of the initial velocity model. An inadequate initial model can lead the tomography inversion to a local minimum, not honouring the correct velocity model. Usually, a good initial velocity model does not present detailed structures but preserves the main velocity trend of the model. 
To generate the initial model for the Deep-Tomography flow, we applied a Gaussian Filter with $\sigma=100$ in both directions and only adjusted the water layer to the original position and values. The model generated is a rough velocity trend of the true model, without any resemblance to the original structures in the subsurface, as represented by the input model of first iteration in Fig.~\ref{fig:tomoflow}. In the absence of the true model, we considered that this velocity trend could be obtained with conventional geophysical methods, like converting a RMS velocity obtained by NMO picking to interval velocity or smoothing an older regional velocity model.

The models that define the input for the next iterations are obtained by reducing the size of the Gaussian filter applied to the true model, using $\sigma=50$ for the second iteration and $\sigma=20$ for the third iteration. The effect of the different filter size over the models are shown in Fig.~\ref{fig:tomoflow} as the input models of the second and third iterations. 
With these smoothing definitions, each iteration defines a degree of model accuracy to be predicted by the DL algorithms.

\begin{figure*}
    \centering
    \includegraphics[width=0.95\linewidth]{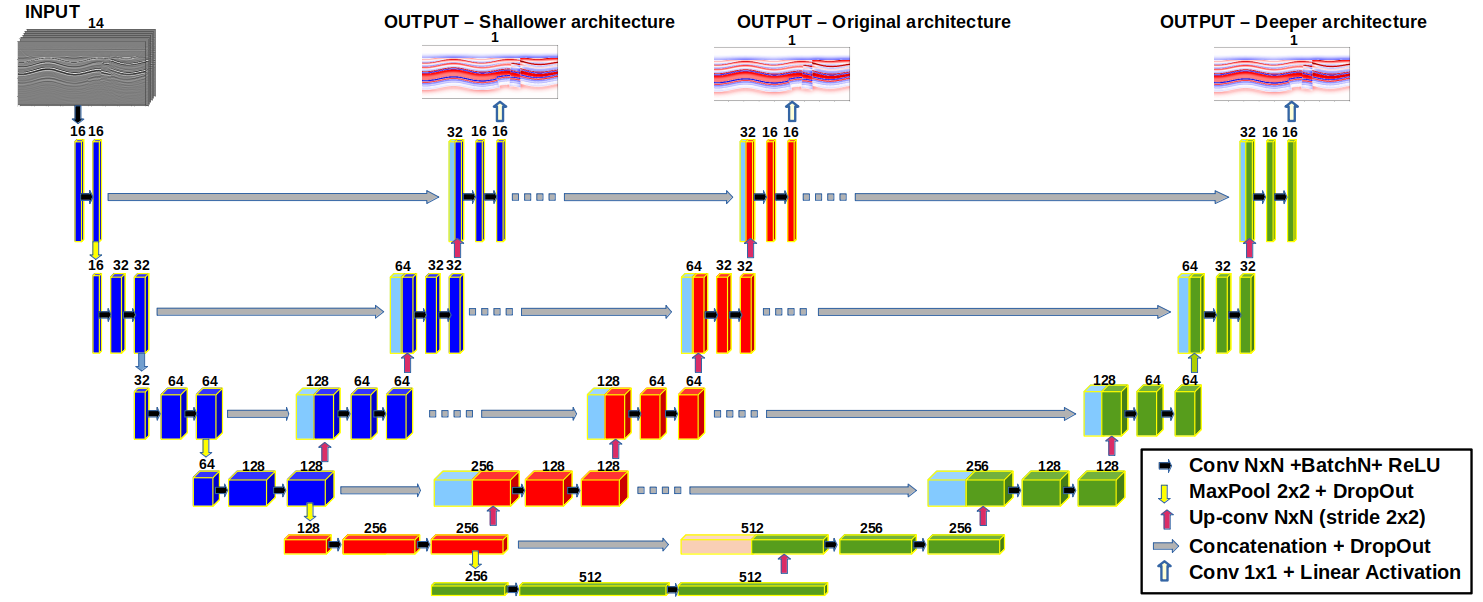}
    \caption{This figure presents a diagram of the three U-Net architectures we used in this work. Blue stands for blocks common to the three architectures in the encoder branch and the blocks of the shallower architecture in the decoder branch. Orange stands for blocks common to original and deeper architectures in the encoder branch and the blocks of the original architecture in the decoder branch. Finally, green stands for blocks exclusively of the deeper architecture. The arrows indicate the operations which transform one block into another, and the numbers over each block indicate the number of channels or features maps.}
    \label{fig:u-net}
\end{figure*}
\section{Deep Learning definitions}
\subsection{Input and labels for training}
The DL algorithm receives common-offset subsurface images, where the network interprets each subsurface offset panel as a channel of a 2-D image. We used fourteen subsurface offsets panels as input. The maximum offset was chosen based on the maximum offset where we observed coherent reflection energy when migrating the data with the initial model. We clipped the amplitude of each subsurface offset panel in percentile 96 and then normalized the amplitude of each panel between 0 and 1.

The expected outputs, i.e, the labels presented during the training process, are the difference between the desired output velocity model for that iteration and the velocity model used to migrate the inputs. Thus, the labels can be interpreted as the corrections applied to the current iteration's velocity model. The labels units are $km/s$. As we want to use an iterative strategy to obtain the true model, the model used in migration and the desired one differs by one degree of smoothing between themselves, as presented in Fig.~\ref{fig:tomoflow}.

With the iterative strategy, it is necessary to train the neural network three times, one time for each iteration. Therefore, we made an independently training process for each iteration, using the correspondent input and outputs for the current iteration to define the network weights. Thus, each network is specialized to make one iteration of Deep-Tomography, which is summarized in Fig.~\ref{fig:tomoflow}.
After the training process, we can use the trained networks to predict an unknown model using only as input the shots acquisition and an initial velocity model like the one in input model of first iteration (Fig.~\ref{fig:tomoflow}). The expected final result of the Deep-Tomography flow is a detailed model, similar to the true model used to model the shots.

As the original dimension of the models and migrated panels are 351 samples vertically and 901 horizontally, it was necessary to adjust these dimensions to fit in the original  U-Net implementation flow. We choose to include one null sample vertically, thus the new size becomes 352, and cut five samples laterally, getting the new horizontal size of 896. Since, when constructing the velocity models, we laterally extrapolated the models to allow the full shot covering inside the interest region, when updating the model iteratively, we also extrapolated the predicted model update. For each training result presented here, we divided the set of 1000 models into training/validation and test. The test set comprises ten per cent of the original set and is the first one separated from the original models, and validation set represents twenty per cent of the remaining models and it is used to evaluate and define the training performance while the test set is applied in final's model performance as a blind test.

\subsection{Network architecture and loss definition}

In this work, the networks investigated to address the task of Deep-Tomography are based on the U-Net architecture \cite[]{UNET}. The U-Net is an encoder-decoder fully convolutional model \cite[]{LONGSHELHAMER} typically used in computer vision related tasks. Encoder-decoder models are often employed in these tasks because they often work with inputs and outputs with the same spatial dimensions, and combine a deep feature extraction branch (encoder) and a precise feature localization branch (decoder) \cite[]{DEEPLABV3}. As an encoder-decoder model, the U-Net has two peculiarities: the shortcut connections between encoder and decoder branches, which reinforces the flow of information in the network; and the symmetric decoder branch, which gradually learns to upsample the feature map produced by the encoder.

Although encoder-decoder models are  commonly used in classification or segmentation tasks, they can be repurposed with an adequate choice of the loss function and normalization of the input and output data. Since our task of generating velocity updates using the U-Net is closer to a non-linear regression problem, we used the mean squared error between predicted and real velocity update as a minimization objective for the training process of our network models.

Since it is a convolutional network, the learning process for the U-Net is driven by weight updates in the convolution kernels after every training epoch. The convolution kernels used in the U-Net design follow the success of earlier convolutional networks \cite[]{DEEP_W_CONV,VGG}, which traded larger kernel sizes for smaller kernels in deeper stacks of convolutional layers to reduce computational cost and avoid overfitting.

In VMB flows, one possible information analysed is the moveout error. Since a moveout error in the migrated CIG indicates the presence of errors in the velocity model. When analysing a moveout error isolated, the only thing that we can state is that the velocity error is above the observed event. However, this error can be localised or spread over a large (lateral/vertical) model region; thus, there is no direct spatial correspondence between the moveout and the velocity error positions. Thus, as our network inputs are migrated panels, we assume or hypothesize that the correct identification of the velocity update (output) goes through the same logic. Thus, the local properties observed in the input data present a long-range connection with the properties of the output data, and a network with short-range connections between different pixels in the input/output images can not produce satisfactory results for our problem.

We chose to represent the long-range dependence by increasing the kernel size. Thus, we can connect distant regions of the migrated panel(input) with the velocity corrections(output). A direct effect of using larger kernels is to increase the number of trainable parameters in the U-Net model. An increase of computing capacity and accuracy of predictions of the network is also expected with the increase of trainable parameters. To distinguish between the effects of a larger kernel and more trainable parameters in the model, we perform comparative experiments using U-Net architectures with a different number of layers: shallow, original, and deep, and with different kernel sizes $K_{size} = [3\times3, 6\times6, 9\times9, 12\times12]$. The architectures with more layers will present more trainable parameters when using the same kernel size. Thus, we can check if a large kernel size is responsible for better performance by evaluating the loss, i.e. the mean squared error, over the validation set for each kernel size investigated.

Fig.~\ref{fig:u-net} shows each design in detail. The shallow model only has three stages in both encoder and decoder branches; the original model, as per \cite[]{UNET}, has 4 staged encoder and decoder branches; and the deeper U-Net has five stages in each branch. Each of these U-Net models is trained with every $K_{size}$ value so that a distinction can be made between the effects produced by different kernel sizes and different model depths.

\subsubsection{Implementation details}

We have included batch normalization and dropout layers in our U-Nets, as shown in Fig.~\ref{fig:u-net}. Dropout \cite[]{DROPOUT} is a simple and effective regularization strategy, while batch normalization \cite[]{BNORM} helps stabilize the training process and make it more efficient by avoiding gradient related problems. All networks were trained with the Adam \cite[]{ADAM} variation of the stochastic gradient descent algorithm, with a base learning rate of $1e^{-4}$ over 100 epochs. One training epoch of the network with the smallest number of parameters takes approximately 15 seconds, and for the network with the largest number of parameters, approximately 110 seconds. All training processes are carried out on one NVIDIA V100 with 32Gb and the implementation was done with the Keras API of the Tensorflow open-source deep learning library.

\section{Results}

\begin{figure}
    \centering
    \includegraphics[width=0.49\linewidth]{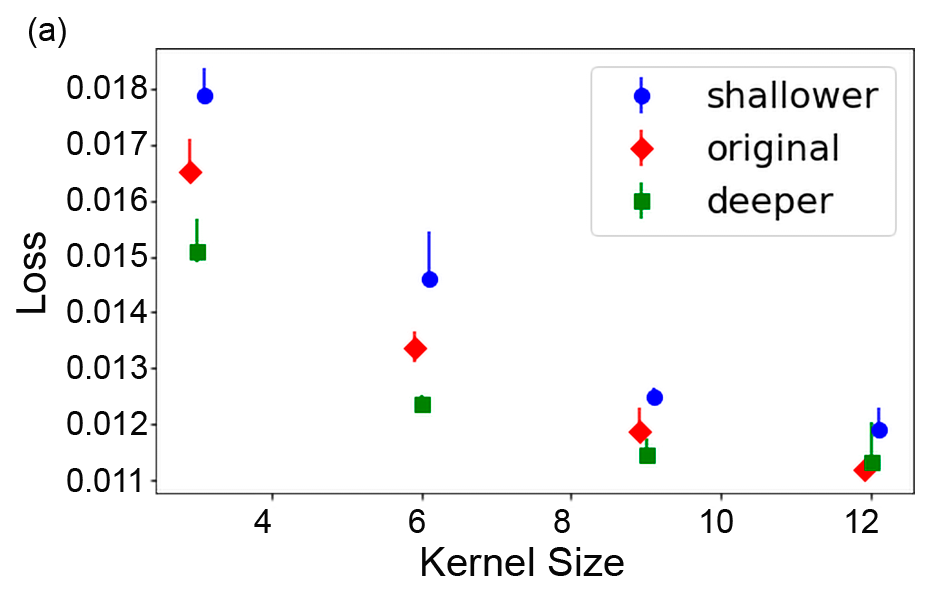}
    \includegraphics[width=0.49\linewidth]{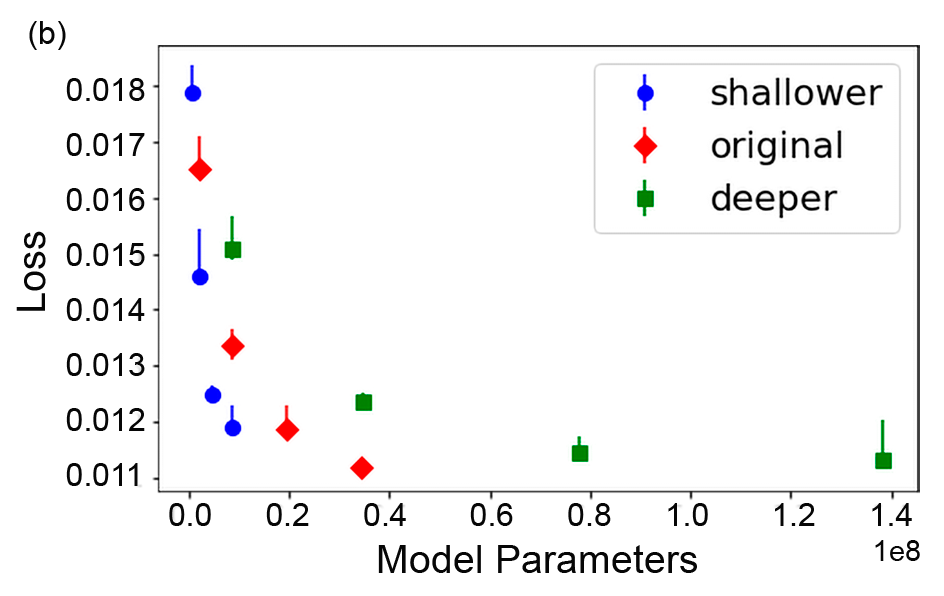}
    \caption{In Fig.~(a) we show the lowest validation loss calculated over the training as a function of the kernel size of convolutional layers in the U-Net. We divided the result into three different architectures, the original U-Net, a shallower, and a deeper version. We can see that larger kernels reduce the loss. Fig.~(b) plotted the same results but as a function of number of trainable parameters of the U-Net.}
    \label{fig:param_und_kernels}
\end{figure}

\subsection{Sensibility to the kernel size and number of trainable parameters}
\label{sec:ks}

This section presents the tests to evaluate the effects of increasing the kernel size. We used $k$-fold cross-validation in the training process \cite{moreno2012study} to account for fluctuations inherent to the random nature of the process, including training sample definition. We shuffled our validation set into five-folds in cross-validation, where each validation set generates a different training process. Our loss measures show the smallest validation loss of the whole process, and the error bars represent the fluctuations of this value over the five training processes.

In Fig.~\ref{fig:param_und_kernels}(a), we show the effect of kernel size over the validation loss for the three proposed U-Net architectures: original, deeper and shallower. We observe that the loss reduces with the increasing of the kernel size for all architectures, and also reduces with the number of layers in the network when analysing the same kernel size. However, the number of layers seems to presents a smaller effect when compared to the kernel size. 

To clarify the effect of kernel size, when compared to the number of trainable parameters of the network, we also plotted in Fig.~\ref{fig:param_und_kernels}(b) the loss as a function of the number of parameters of the U-Net model. The plot with the number of parameters shows that despite the shallower U-Net having a reduced number of parameters, it reaches a result as good as the deeper U-Net with the largest kernel size, which confirms the expected positive effects of increasing kernel size over the prediction accuracy. The results presents in Fig.~\ref{fig:param_und_kernels} accounts for the training of U-Net of the third iteration. The loss results presented here were obtained for the third iteration. The same effect is observed for the other iterations.

\begin{figure*}
\includegraphics[scale=0.33,trim={6.5cm 6.8cm 3.5cm 6.8cm}]{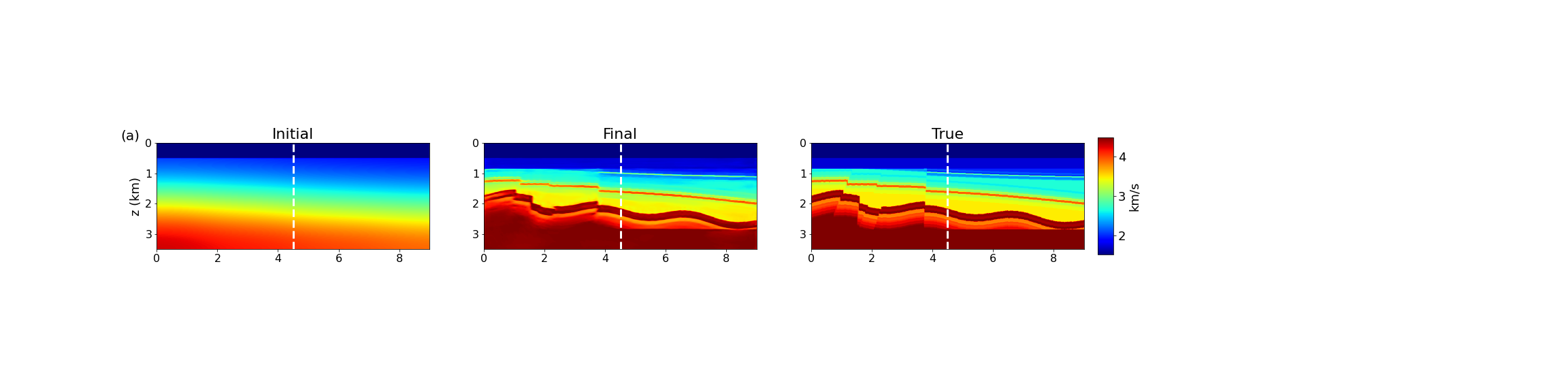}
\includegraphics[scale=0.33,trim={6.5cm 6.8cm 3.5cm 6.8cm}]{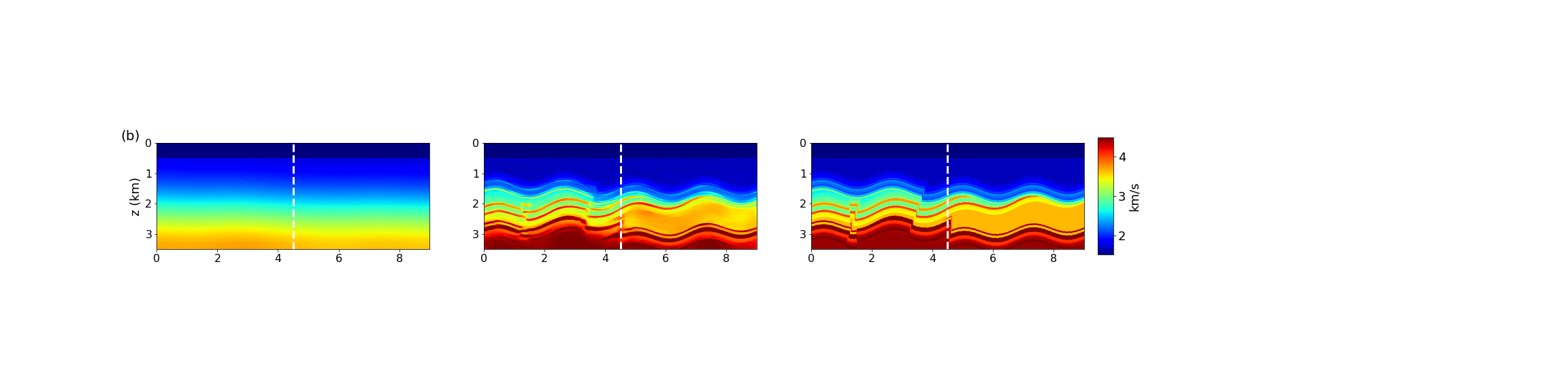}
\includegraphics[scale=0.33,trim={6.5cm 6.8cm 3.5cm 6.8cm}]{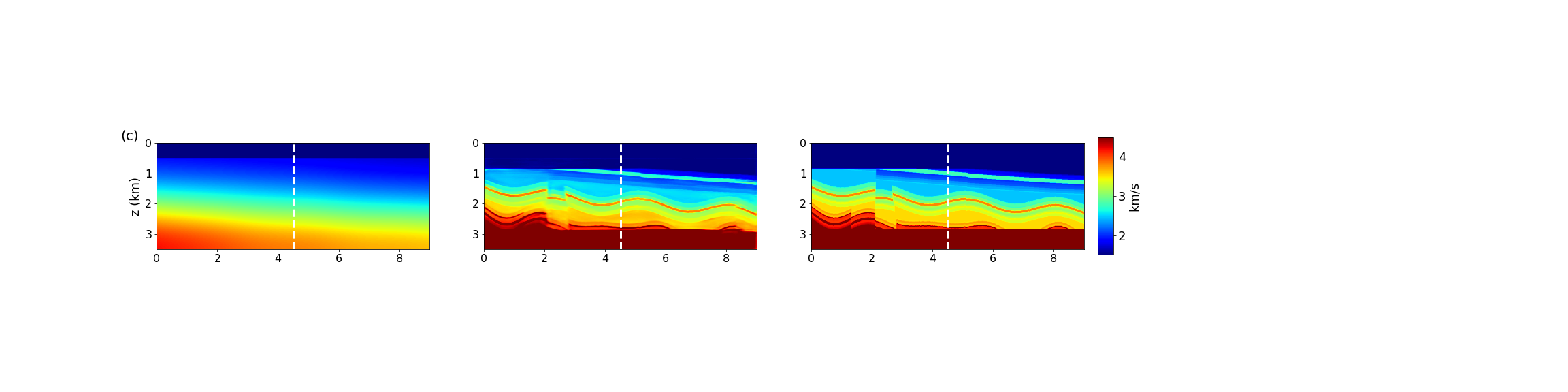}
\includegraphics[scale=0.33,trim={6.5cm 6.8cm 3.5cm 6.8cm}]{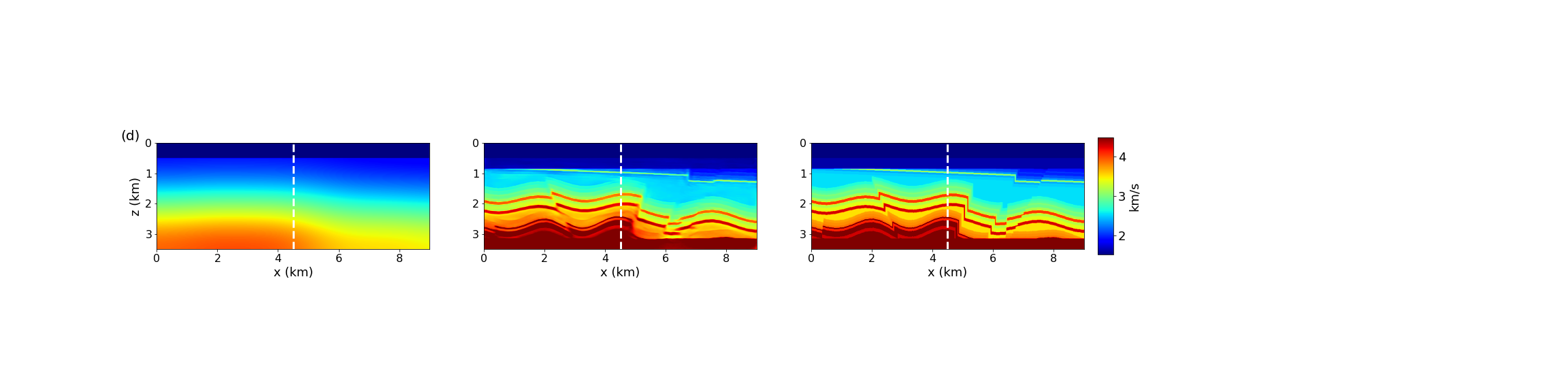}
    \caption{These figures show models extracted from test set and their response to Deep-tomography. The left column shows the initial models which started the iterative process, the middle column the final results after three iterations, and the right column the true models. The white vertical line marks the position where we extracted velocity profiles plotted that are shown in Fig.~\ref{fig:log_evolution}.}
    \label{fig:model_evolution}
\end{figure*}
\begin{figure*}
\includegraphics[width=0.99\textwidth]{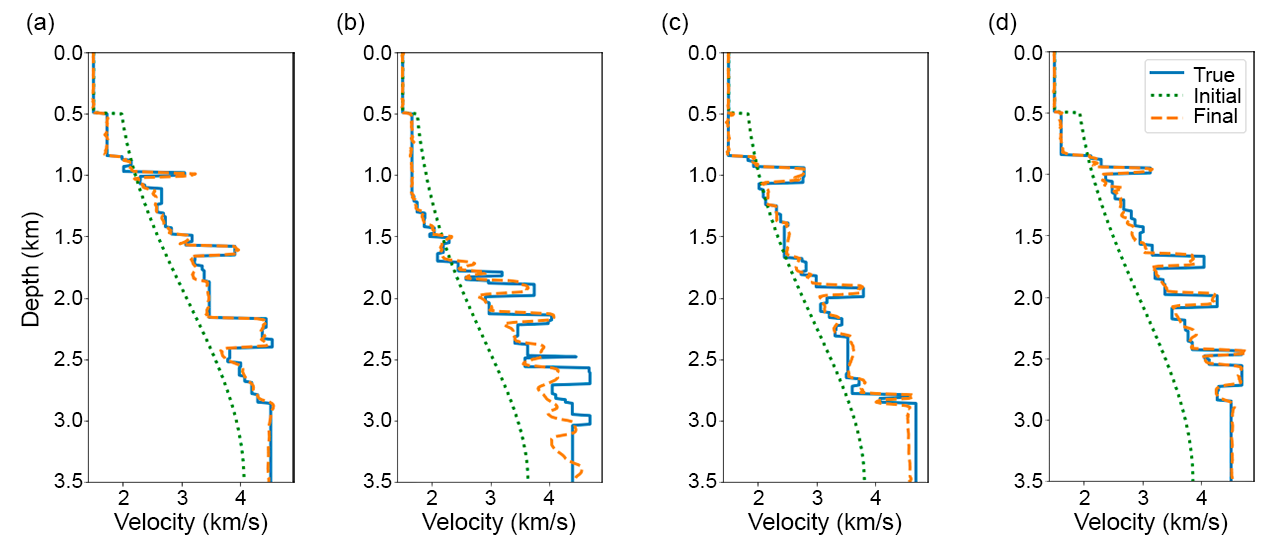}
\caption{Velocity profiles extracted from the models presented in Fig.~\ref{fig:model_evolution}.}
    \label{fig:log_evolution}
\end{figure*}

\subsection{Deep-Tomography results over the test set}

As previously mentioned, from our original set of $1000$ models, we separated ten per cent of the models for the test set. The test set was not used in the training and validation process of the network and is used only to perform a blind and independent evaluation of the prediction ability of the method.
Fig.~\ref{fig:model_evolution} presents the result of the Deep-Tomography for four models from the test set. As mentioned before, Deep-Tomography starts migrating the shots with a smoothed model that we plot in the first column of Fig.~\ref{fig:model_evolution}. 
We update this initial model in a cascade of processes, which consists of predicting the model update with the U-Net for the current iteration, migrating with the updated model, and them predicting in the next iteration until reaching the final model, as was previously represented in Fig.~\ref{fig:tomoflow}.
After three iterations of model updating, we plot the final model in the middle column of Fig.~\ref{fig:model_evolution}, and for comparison, the true model in the right column. All the results use a shallower U-Net with kernel size equal to twelve.

It is possible to see in Fig.~\ref{fig:model_evolution} that Deep-Tomography predicts the velocity models with a high resemblance with the true model, even for the small structures and thin layers. The faults in the models were also preserved but not with the same detail level presented in the true model. 
A critical aspect of being considered here is that the models accumulate mistakes in predictions when they pass from one iteration to the next. However, the U-Nets seem to be able to correct them. Therefore, the final error in prediction is smaller than the sum of the errors measured along with the iterations.

A notable feature is that the quality of the predicted model does not degrade significantly with increasing depth. Usually, when making a VMB with conventional methods, the best results are achieved when building the model by layers \cite[e.g.,][]{LS2,LS1}. However, Fig.~\ref{fig:model_evolution} suggests that, when using Deep-Tomography, at least for the maximum depth investigated here, the layer striping practice could not be necessary. 
In Fig.~\ref{fig:model_evolution} a white vertical trace indicates the position where we extracted a vertical profile from the models for better comparisons. The vertical profiles are in Fig.~\ref{fig:log_evolution}, and they correspond in order to the models presented in Fig.~\ref{fig:model_evolution}. For a better understanding of the degree of complexity of the task, we plotted the initial, final, and true velocity models. 
In almost all cases, the profile signals that the final velocity, predicted by the trained U-Nets, is close to the correct velocity, although it is not a perfect match.
The second model is a good example where prediction presents a more significant amount of errors. When we check the position of the profile in the second model in Fig.~\ref{fig:model_evolution}, it is possible to see that it crosses a fault region, which is probably the cause of the highest errors.

A crucial aspect regarding predictions here is that the high level of accuracy is only possible with the iterative strategy. Initially, we tried an approach of building the model using only one iteration, which has as output label the difference between the input model for iteration first and the real model.
The network made poor predictions in one step. The possible cause of this result is that when we migrate data with a model that is very far from the true model, there are few focused events of faults, thin layers, and other details that our complex set of models presents. As a result, a limited amount of reflection energy correlates at this level in the migrated image that feeds the prediction flow. If our true models were simpler, i.e., with fewer structures and details, a reduced number of iterations would be enough to predict an accurate model.

\begin{figure}
\includegraphics[scale=0.57,trim={0.5cm 2cm 0cm 2cm}]{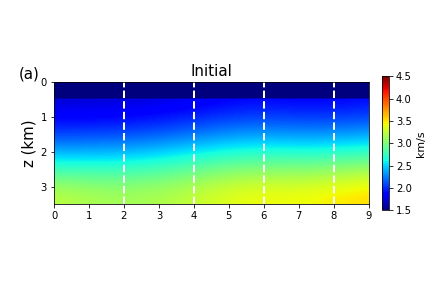}\\
\includegraphics[scale=0.57,trim={0.5cm 2cm 0cm 2cm}]{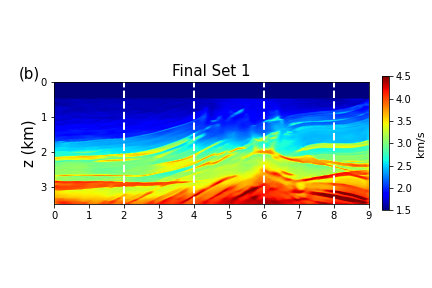}\\
\includegraphics[scale=0.57,trim={0.5cm 2cm 0cm 2cm}]{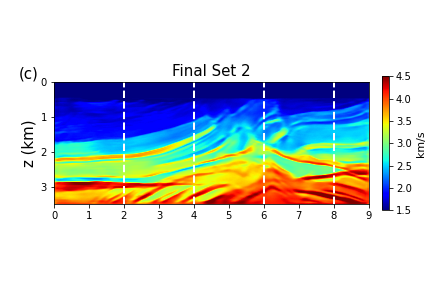}\\
\includegraphics[scale=0.57,trim={0.5cm 2cm 0cm 2cm}]{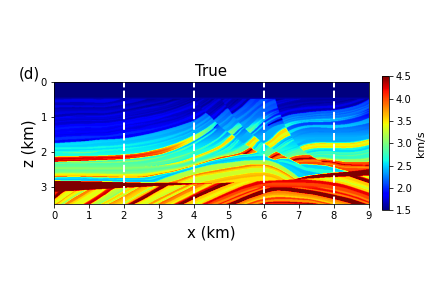}\\
\caption{Fig.~(a) shows the initial model used as input for Deep-Tomography of Marmousi model, 
Fig.~(b) the final model obtained using Set 1 as training model for U-Nets, Fig.~(c) the final model obtained using Set 2 as training model for U-Nets, and Fig.~(d) is the true model used in shots simulation. The white vertical lines indicate the positions where we extracted the velocity profiles showed in Fig.~\ref{fig:log_marmousi}.}
\label{fig:model_marmousi}
\end{figure}

\begin{figure}
\includegraphics[height=0.60\textwidth]{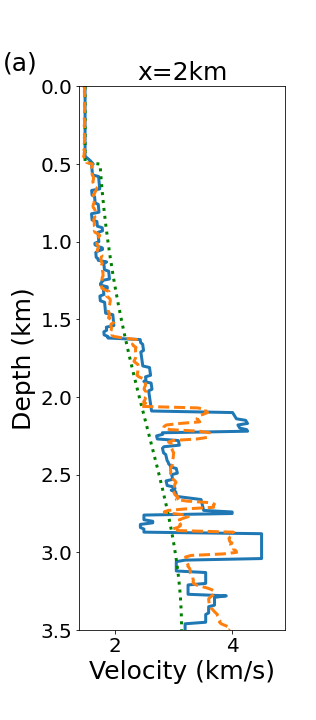}
\includegraphics[height=0.60\textwidth]{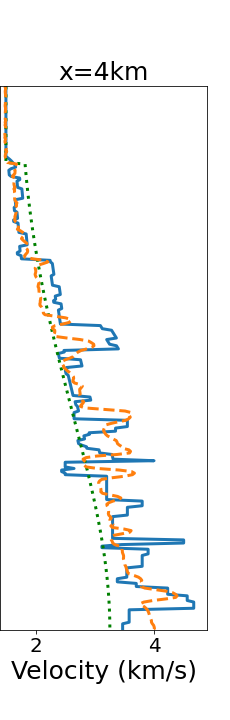}
\includegraphics[height=0.60\textwidth]{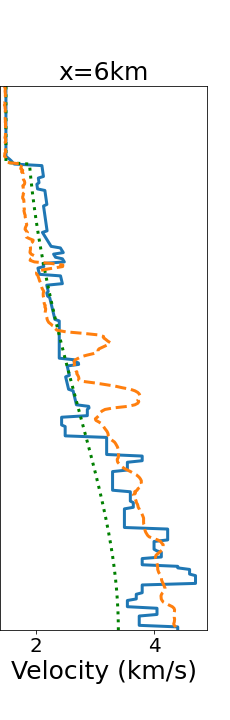}
\includegraphics[height=0.60\textwidth]{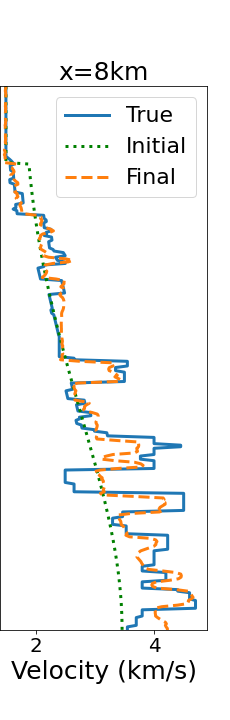}\\
\includegraphics[height=0.60\textwidth]{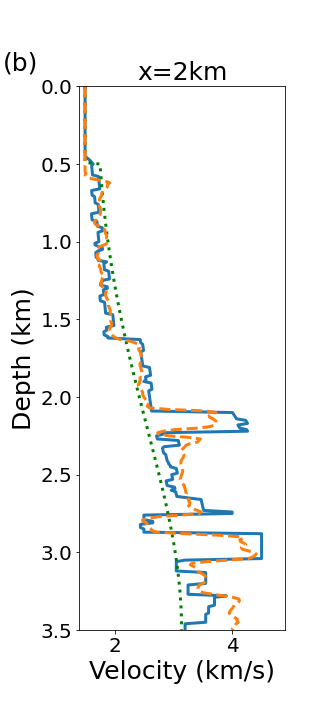}
\includegraphics[height=0.60\textwidth]{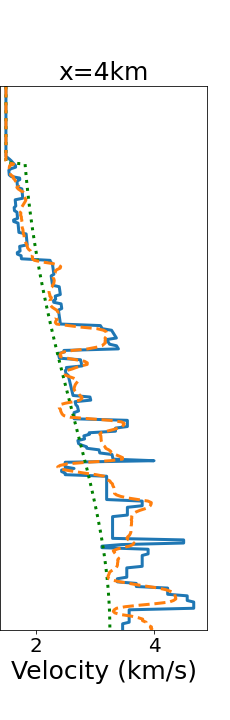}
\includegraphics[height=0.60\textwidth]{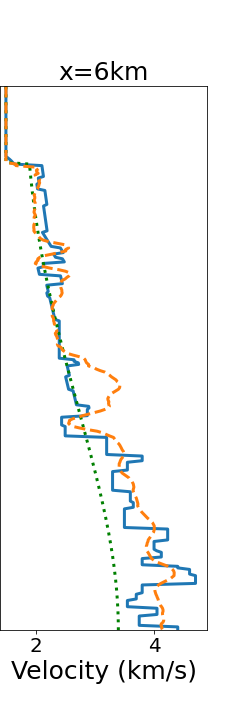}
\includegraphics[height=0.60\textwidth]{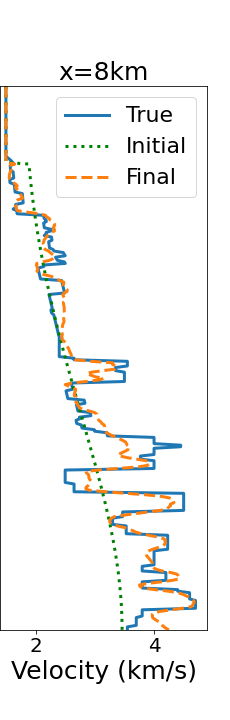}\\
    \caption{Fig.~(a) shows the velocity profiles extracted from the model in Figs.~\ref{fig:model_marmousi}(b) and Fig.~(b) the velocity profiles extracted from the model in Fig.~\ref{fig:model_marmousi}(c). Both present for reference the velocity profiles of the initial model(Fig.~\ref{fig:model_marmousi}(a)) and for the true model(Fig.~\ref{fig:model_marmousi}(d)).}
    \label{fig:log_marmousi}
\end{figure}

\begin{figure*}
    \centering
    \includegraphics[scale=0.30,trim={1cm 1cm 1cm 1cm}]{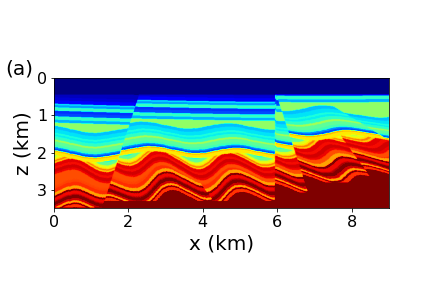}
    \includegraphics[scale=0.30,trim={1cm 1cm 1cm 1cm}]{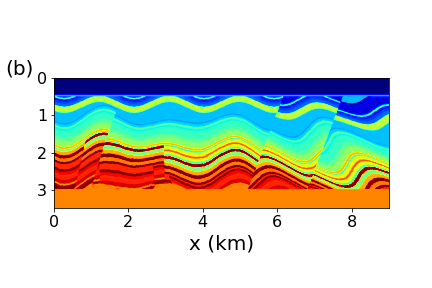}
    \includegraphics[scale=0.30,trim={1cm 1cm 1cm 1cm}]{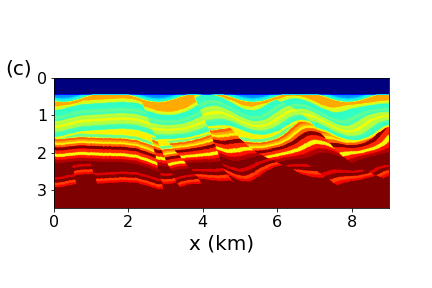}
    \includegraphics[scale=0.312,trim={1cm 1cm 1cm 1cm}]{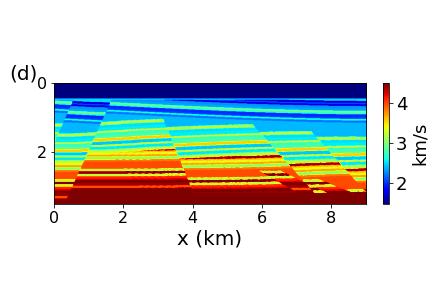}
    \caption{Examples of models from training Set 2 to illustrate the modifications done to predict a better result for Marmousi model. In Fig.~(a) and (b) we show examples with a low/high velocity layer contrasting with the surroundings, Fig.~(c) shows examples of listric faults and, Fig.~(d) illustrate faults with low dips.}
    \label{fig:set2}
\end{figure*}

\subsection{Results for the Marmousi model}
In order to evaluate the generalization of Deep-Tomography, we performed a series of tests over Marmousi data-set \cite[]{marm}. Marmousi is a complex structural model and is a widely used benchmark to evaluate imaging and velocity model building techniques. Fig.~\ref{fig:model_marmousi}(d) shows a plot of Marmousi model.

Initially we used the U-Nets trained with the original data-set proposed in \ref{sec:models}, here named as Set 1. The U-Net architecture chosen in this case was the original with kernel size equal to twelve. The initial model used in the flow is plotted in Fig.~\ref{fig:model_marmousi}(a), and the final prediction obtained is plotted in Fig.~\ref{fig:model_marmousi}(b), with the correspondent velocity profiles in Fig.~\ref{fig:log_marmousi}(a). Despite a considerable increasing of complexity when compared with the initial model, we observed that the predicted model did not account for high contrast layers, specially the ones which presents low-velocity, what can be confirmed by analysing the velocity profiles in Fig.~\ref{fig:log_marmousi}(a).
Besides, it presented poor predictions in the central portion of the model, specially in recovering appropriately the faults, even for the shallow zone. Inaccuracies in this zone are partially expected since this is the most structural challenge portion of the model, but even the velocity trend is misrepresented. Thus we investigated further improvements in our method to better recover the Marmousi model.

Representativeness in the training set is a critical point in machine learning by construction, since the network learns from what it is presented to it. Therefore, a training dominated by extremes and anomaly cases are likely to have a poor performance. However, the true parameter space probability distribution of real data and the relevant parameters to describe the data are unknown. For example, the relevant parameters could be the geometry of the layers, the velocity range, the reflectivity. Thus, we add additional steps in the geologically flow that built the training set, increasing its complexity to check if the new trained U-Nets obtain better results with the Marmousi model. \\

When comparing the velocity trends of our training set 1 and the Marmousi model, we noted that Marmousi presents some high contrast low-velocity layers, absent in training set 1. Set 1 only presents high-velocity layers, as described in the data-set construction section. Moreover, we adjusted a minimal distance between the water layer and the first sediment interface when defining the depth of the final depositions in set 1. Thus, we constructed a new training data-set, including low-velocity layers. We also reduced the water layer's distance to the first sediment interface to account for the shallow faults that dominate the central portion of the model. Besides, we improved the creation of faults in the model that previously presented a stepped shape, as can be seen from examples in Figures \ref{fig:models} (c) and (d). In the improved flow, we are creating faults that are more similar to the ones observed in the Marmousi model, without the steps, with lower dips and with the linear or listric form. In Figure \ref{fig:set2}, we show some examples of this new training set that we named Set2. 

Then, we used Set 2 to train the U-Nets and evaluate the effect of these simple modifications over Marmousi's predictions. Fig.~\ref{fig:model_marmousi}(c) shows the predicted model with the new training set. It is possible to observe that the predicted model represents better the True model when compared with the past result (Fig.~\ref{fig:model_marmousi}(c)), which can be confirmed by analysing its velocity profiles (Fig.~\ref{fig:log_marmousi}(b)). Furthermore, from a smooth and non-structured model, we can see that the Deep-Tomography flow recovered the significant layers of the true model. 
The most challenging prediction points are in the central faulted region of the model, confirmed by the velocity profiles, where the predictions between the horizontal position of 5km and 7km present the worst accuracy. Another practice adopted in the flow for the Marmousi model was to zero the model updates inside the water layer, and to clip anomalous values restricting velocity values between 1500m/s and 4700m/s.

\begin{figure}
\begin{center}
\includegraphics[scale=0.44,trim={1cm 0cm 1cm 1
0cm}]{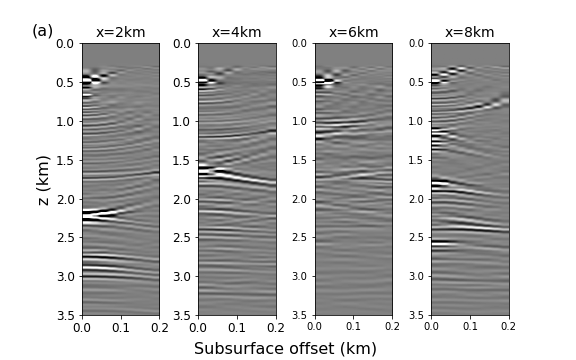}
\includegraphics[scale=0.44,trim={1cm 0cm 1cm 0cm}]{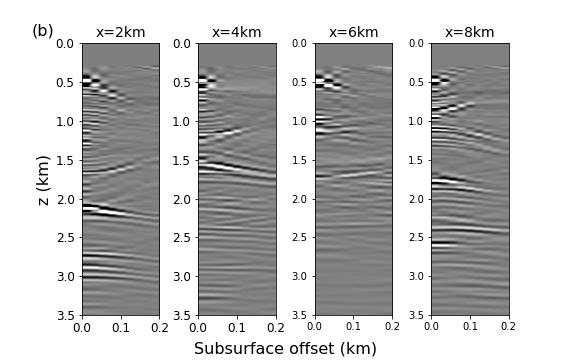}
\includegraphics[scale=0.44,trim={1cm 0cm 1cm 1
0cm}]{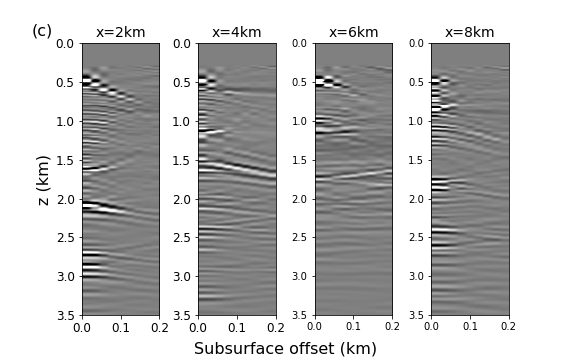}
\includegraphics[scale=0.44,trim={1cm 0cm 1cm 1
0cm}]{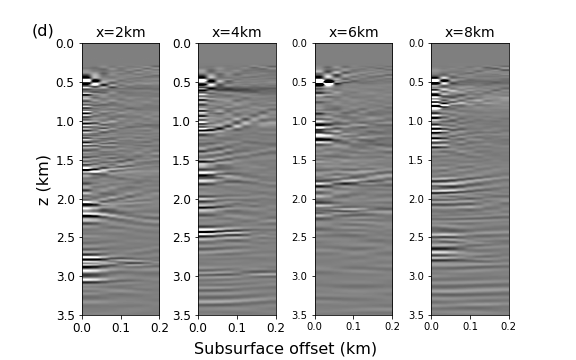}
\includegraphics[scale=0.44,trim={1cm 0cm 1cm 1
0cm}]{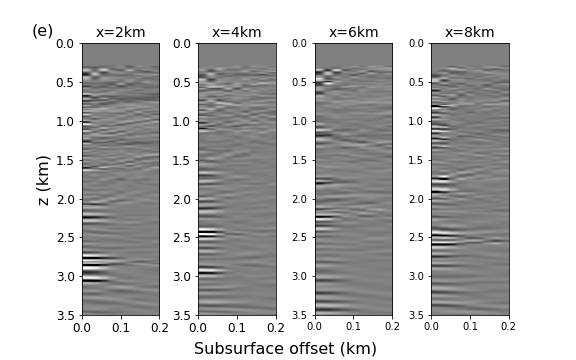}
    \caption{Focalization test for the flow which obtained the Marmousi model with Set 2. We selected some positions to show the common image gathers for the subsurface offset migrated with the initial model (Fig.~(a)), model obtained with the first iteration (Fig.~(b)), model obtained with the second iteration (Fig.~(c)), model obtained with the final iteration (Fig.~(d)) and, migrated with the real model (Fig.~(e)).}
\end{center}
    \label{fig:cig_marmousi}
\end{figure}

\begin{figure*}
\begin{center}
\includegraphics[scale=0.43,trim={6.5cm 6.6cm 3.5cm 6.6cm}]{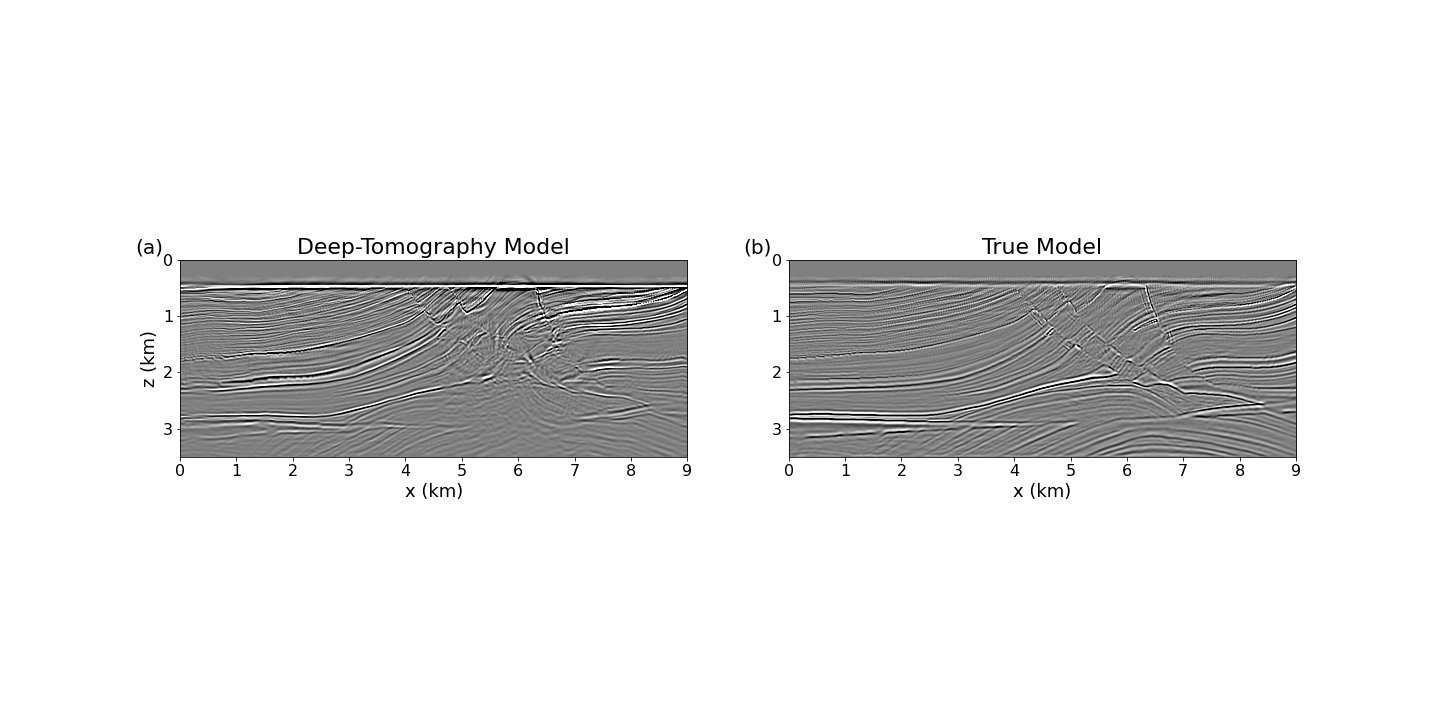}\\
    \caption{Fig.~(a) shows the migrated image of Marmousi using the model predicted with Deep-Tomography(Fig.~\ref{fig:model_marmousi}(c)) and Fig.~(b) the image generated using the true Marmousi model(Fig.~\ref{fig:model_marmousi}(d)).}
\end{center}
    \label{fig:mig_marmousi}
\end{figure*}

In order to compare the results obtained with Set1 and Set2, we measured the mean absolute error, and the structural similarity index(SSIM) \cite[]{SSIM} between the predicted model and the true Marmousi model. The mean absolute error for Set 1 is equal $235m/s$ and for Set 2 is equal to $185m/s$.  SSMI for Set 1 is equal to $0.59$ and for Set 2 is equal to $0.68$. The similarity index is defined inside the range from 0, as the worst score, to 1, as the best score. Thus, both metrics confirm that using Set 2 we obtained a better final prediction.

Figure \ref{fig:cig_marmousi} shows the evolution of the common image gathers(CIG) migrated with the velocity models obtained during the flow implemented with Set 2. Since we are analysing the subsurface offset, the reflections are expected to focus around zero offset when improving the velocity model. Fig.~\ref{fig:cig_marmousi}(a) shows the CIG generate with the input model, Fig.~\ref{fig:cig_marmousi}(b) with the model after the first iteration, Fig.~\ref{fig:cig_marmousi}(c) with the model after the second iteration and, Fig.~\ref{fig:cig_marmousi}(d) with the final model. It is possible to observe that, in general, focalisation increases progressively with the iterations. In Fig.~\ref{fig:cig_marmousi}(e) we show the CIG migrated with the real model for reference. Another possible evaluation of a velocity model quality is to migrate the shots with this model and see how are the structures' positions and continuity. For example, in Figure \ref{fig:mig_marmousi} we show the migration with the predicted model from the Deep-Tomography and the true model. As expected, we observe the best results outside the central faulted region. However, the predicted model imaged satisfactory the shallow structures in the central portion of the model and the significant structures in the deep portions.

\begin{figure}
\includegraphics[scale=0.57,trim={0.0cm 1.8cm 0cm 2cm}]{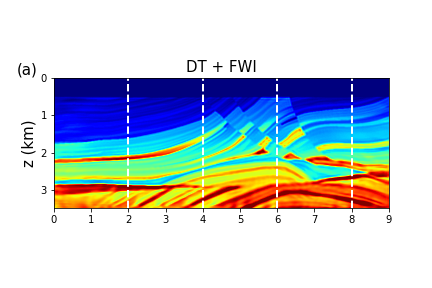}
\includegraphics[scale=0.60,trim={0.5cm 2cm 0cm 2cm}]{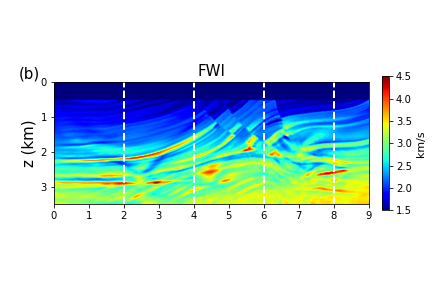}\\
\caption{Fig.~(a) show the velocity models obtained with FWI using as the initial model the final result of Deep-Tomography (Fig.~\ref{fig:model_marmousi}(c)) and, Fig.~(b) the model obtained with FWI using as the initial model the input of Deep-Tomography flow (Fig.~\ref{fig:model_marmousi}(a)). The white vertical lines indicate the positions where we extracted the velocity profiles showed in Fig.~\ref{fig:log_fwi}.}
\label{fig:mod_fwi}
\end{figure}

\begin{figure}
\includegraphics[height=0.60\textwidth]{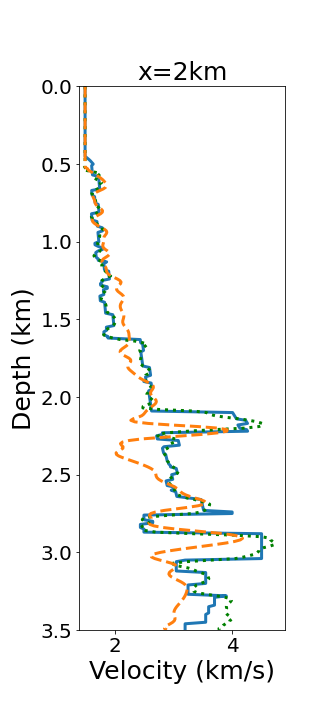}
\includegraphics[height=0.60\textwidth]{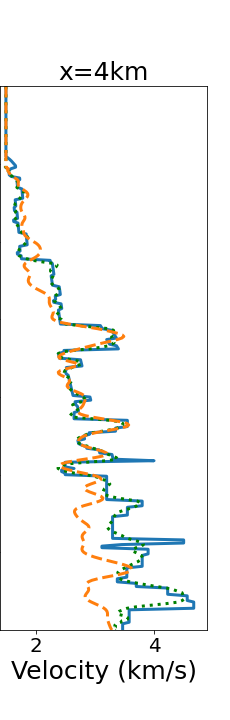}
\includegraphics[height=0.60\textwidth]{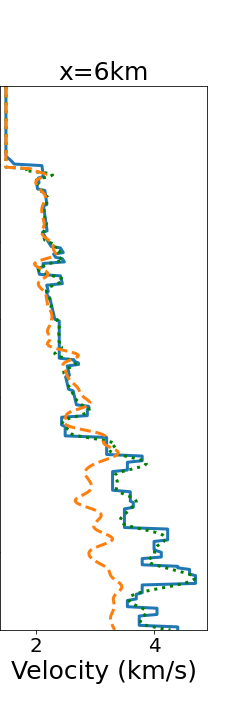}
\includegraphics[height=0.60\textwidth]{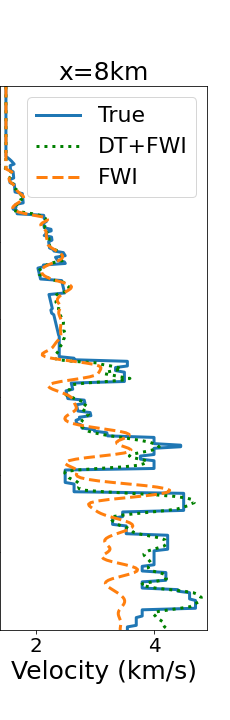}\\
\caption{Velocity profiles extracted from models presented in Figure \ref{fig:mod_fwi}(a) (DT+FWI) and  Figure \ref{fig:mod_fwi}(b)(FWI). We also plotted the vertical profile of the true Marmousi model presented in Figure \ref{fig:model_marmousi}(d) for reference of the expected result.}
\label{fig:log_fwi}
\end{figure}

\subsection{Comparison and combination of Deep-Tomography with conventional methods}

Some open questions concerning the use of DL for VMB are the performance comparison with conventional methods and if it is possible to implement a mixed flow which combines conventional and DL methods. Combining methods may cover the limitations of both approaches, resulting in a more reliable velocity model. In this section, we applied a FWI algorithm using two different initial models: the final model obtained with Deep-tomography (Fig.~\ref{fig:model_marmousi}(c)), which we named as DT+FWI result, and the initial model used in Deep-Tomography flow (Fig.~\ref{fig:model_marmousi}(a)), which we named FWI result. Thus, we intended to compare flows that started with the same model, but using different methods. The shots peak frequency was reduced in both tests, with the peak at 8Hz to avoid artefacts that could appear at FWI inversion. We also changed the receiver distributions to a fixed-spread configuration covering all the surface in order to get longer offsets and increase FWI performance.
The optimization algorithm used was the gradient descent preconditioned by the inverse of pseudo-hessian \cite[]{shin2001efficient}. The velocity model update step is defined by a Wolfe type line-search with backtracking with a maximum step-length of 100~m/s \cite[]{wright1999numerical}. 
The velocity models presented here are the results after two hundred iterations; however, the DT+FWI result reaches a stable misfit after approximately fifty iterations. 

Figure \ref{fig:mod_fwi}(a) shows DT+FWI result. It is possible to observe that structures are recovered with great detail, even in the deeper portions of the model. To better compare the result with the real Marmousi model, we plot in Figure \ref{fig:log_fwi} the profiles for the positions signalized with the vertical lines in models. Figure \ref{fig:mod_fwi}(b) shows the FWI result. The profiles of this model are also plotted in Figure \ref{fig:log_fwi}. The result obtained using only FWI reached a high degree of accuracy in the shallow portions of the model; however, it presents a high mismatch with the real model in the deeper portions. This behaviour is not observed in the result starting from the model obtained with Deep-Tomography. Besides, the convergence for DT+FWI case is faster.

\section{Discussions}

This paper presents a new DL method to infer the velocity model from seismic shots. Our prediction flow, which we named
as Deep-Tomography, does not solve the model in one step but breaks the process into iterations. Each iteration increases the accuracy of the velocity model progressively until converging to the true model. Despite being well-established method in VMB flows, this iterative approach was not previously used associated with DL to obtain the velocity models.
The proposed flow accurately predicted the test set from an unstructured initial model using only three iterations. The significant differences between the predicted and true models occur majorly over the faults.

We chose to use the migrated common subsurface offset panels as input for training/prediction processes. Despite being an uncommon choice in conventional geophysical methods, the common subsurface offsets proved to be a good source of information about the velocity model and present some advantages compared with angle gathers; the generation of such data is simple, and it reduces the amount of information(channels) for the training and predicting process. The migration result is used directly, without the necessity of moveout picking. With further studies, it can be possible to test a cheaper migration to generate the input data, like phase shift migrations, especially for the areas where the velocity models are structurally simple.  

With our iterative approach, the total amount of processing computing increases when compared with the previously one-step proposed methods of VMB with DL. Therefore, at least three migrations would be necessary until reaching an approximated final model. However, the Deep-Tomography reaches an accurate and high resolution model with a reduced number of iterations even for structurally complex models. Such level of resolution is partially reached because the U-Net was trained to predict labels with high resolution, an alternative choice would be to reach a final model with a lower level of resolution, more compatible with the ones obtained with MVA methods. Moreover, the high resolution level we obtained is very hard to recover using only conventional methods, which would be at least bounded by the high frequency limit of the seismic data. The higher resolution was intentionally chosen to show the potential of the method when using a representative training set. 

The network architecture used to solve the Deep-Tomography problem was the U-Net. We investigated, for the purposed problem, the effects of changing the convolutional kernel's sizes, which partially solves the long-range dependency between velocity errors and the correspondent events observed in the migrated data.
A possible alternative to accounting in the network architecture for long-range dependence between structures in output and input  could be the use of recurrent models that has the potential of keeping information from the velocity errors presented in the layers above a determined depth \cite[]{med1,med2}. One important aspect is that recurrence usually increases the training complexity and the use of GPU memory, thus we chose to represent the long-range dependence without recurrence by increasing the kernel size of a simple U-Net. 
We also investigated reducing the number of layers in U-Net, since a shallower network is easier and lighter to train and presents almost the same loss as a deeper version for a selected kernel size. The shallower version proved to present equivalent results over the test set, however in the hard cases, like in the Marmousi benchmark model, the original architecture still presents the best results.

When we tested Deep-Tomography over the Marmousi model, some simple modifications in the training set were necessary to reach a good result. The model predicted for the Marmousi data-set is surprisingly structured and detailed for a method that uses only input data analogous to Migration Velocity Analysis (MVA) methods. Generalization is a known bottleneck in DL methods; the power of the methods is bounded by the appropriate construction of the training set, which has to be representative to the features that the trained network will predict. One important question is how to capture the essential features and include them in the training set. Thus, we must carefully define the training set models to represent the main geological features observed in the studied region. Also for the Marmousi model, we performed a comparison with FWI and a combined flow using the model obtained with Deep-Tomography as the initial model for FWI. The results obtained with the combined flow recovers with high fidelity the model structures, correcting the mismatch of the the small structures in the central portions where Deep-tomography failed. Compared with FWI which used as input a smooth model, the combined flow converged faster and with better results particularly for the deeper regions.

To apply the proposed technique to real data would require some adaptations. The first one is the choice of a neural network architecture that deals with 3D data, some works can guide this choice \cite[]{3dunet1,3dunet2,med1,med2} in order to capture the accurate model update over a 3D framework. An architecture which deals with 3D data requires an intensive use of computational resources, such as memory and processing capacity. To overcome this difficult, the data should be split in patches, updating the model over this patches, and them concatenating  the full model in an adequate way. Besides the mentioned adaptations related to geometry of a real data, synthetic training data-set for the real case must represent effects related to noise and anisotropy in simulations; if the phase and amplitude spectrum of the source wavelet in field data vary from the one used in the training process and, it should be necessary to make these wavelets similar before the migration by applying a matching filter, as was made by \cite[]{paper4}.
Another point of attention when constructing the training data-set is to represent appropriately the structures and velocity ranges expected to be found in the real model, which should take into account the geology of the region to be imaged.

\section{Conclusions}
This work uses the U-Net to determine the velocity model iteratively in a flow similar to the one used in Tomography. We trained a U-Net to perform the iterations using migrated data, starting from a simple smooth model and progressively increasing the model resolution, in a process we called Deep-Tomography. To validate the idea, we defined a flow that constructed a synthetic training/test data set with high geological complexity.  
We proposed a modification in the conventional U-Net architecture, which increased the kernel size of the convolutional layers and proved to be efficient to reduce the loss of predictions. The large kernel accounts for the long-distance effect that a velocity error has over the moveout events in a migrated image.\\
Over the test set, the Deep-Tomography flow predicted the velocity models with high quality. When testing the proposed method over the Marmousi benchmark, the prediction are visually inaccurate, not representing well even the shallow faults. Then, we tested a new flow to define the models used as the training set of U-Net, generating a new velocity distribution with spikes of low and high velocity, and defining the faults with more geological fidelity. After these modifications, the evaluated predictions scores increased significantly. The final predicted model reached a reasonable level, especially when considering the Marmousi's model complexity. We tested a hybrid flow, using as input for FWI the model obtained with Deep-tomography, obtaining a final model with a high degree of similarity to the actual model.\\

\section{Acknowledgements}
We thank the editor Herve Chauris, assistant editor Louise Alexander and one anonymous reviewer for the constructive suggestions and comments that improved the quality of this work. APOM thanks Petrobras for sponsoring her postdoctoral research and for the permission to publish this work. JCC acknowledges the CNPq financial support through the INCT-GP and the grant 312078/2018-8 and Petrobras. CRB acknowledges the financial support from CNPq (316072/2021-4) and FAPERJ (grants 201.456/2022 and 210.330/2022). Finally, the authors acknowledge the LITCOMP/COTEC/CBPF multi-GPU development team for supporting the Artificial Intelligence infrastructure and Sci-Mind’s High-Performance multi-GPU system, and to SENAI CIMATEC Supercomputing Center for Industrial Innovation, for the cooperation, supply and operation of computing facilities.

\section{Data availability}
Data associated with this research are confidential and cannot be released.


\bibliographystyle{gji.bst} 
\bibliography{v0_ref}
\end{document}